\documentclass[useAMS,usenatbib]{mn2e}
\usepackage{graphics}
\usepackage{epsfig}
\usepackage{graphicx}

\newcommand{\apj}{ApJ}
\newcommand{\apjs}{ApJS}
\newcommand{\mnras}{MNRAS}

\newcommand{\aap}{A\&A}
\newcommand{\araa}{ARA\&A}
\newcommand{\apjl}{ApJL}
\newcommand{\aj}{AJ}
\newcommand{\nat}{Nature}

\newcommand{\mbh}{M_{\rm bh}}
\def\ltsima{$\; \buildrel < \over \sim \;$}
\def\simlt{\lower.5ex\hbox{\ltsima}}
\def\gtsima{$\; \buildrel > \over \sim \;$}
\def\simgt{\lower.5ex\hbox{\gtsima}}

\newcommand\ledd{{L}_{\rm Edd}}

\def\msun{{\,{\rm M}_\odot}}

\def\del#1{{}}

\title[Outflows of stars due to quasar feedback]{Outflows of stars due to quasar feedback}

\author[K. Zubovas, S. Nayakshin, S. Sazonov, R. Sunyaev]{Kastytis
  Zubovas$^{1}$, Sergei Nayakshin$^{1}$, Sergey Sazonov$^{2,3}$ and Rashid
  Sunyaev$^{3,2}$\\
$^{1}$ Department of Physics \& Astronomy, University of Leicester, Leicester, LE1 7RH, UK\\
$^{2}$ Space Research Institute, Russian Academy of Sciences, Profsoyuznaya 84/32, Moscow 117997, Russia \\
$^{3}$ Max-Planck-Institut f\"{u}r Astrophysik, Karl-Schwarzschild-Str. 1, Garching 85741, Germany \\
{E-mail:~} {\rm Kastytis.Zubovas@le.ac.uk}}

\begin{document}

\date{Received}
\pagerange{\pageref{firstpage}--\pageref{lastpage}} \pubyear{2012}
\maketitle
\label{firstpage}

\maketitle

\begin{abstract}
Quasar feedback outflows are commonly invoked to drive gas out of galaxies in
the early gas-rich epoch to terminate growth of galaxies. Here we present
simulations that show that AGN feedback may drive not only gas but also stars
out of their host galaxies under certain conditions. The mechanics of this
process is as following: (1) AGN-driven outflows accelerate and compress gas
filling the host galaxy; (2) the accelerated dense shells become
gravitationally unstable and form stars on radial trajectories.  For the
spherically symmetric intial conditions explored here, the black hole needs to
exceed the host's $M_\sigma$ mass by a factor of a few to accelerate the
shells and the new stars to escape velocities. We discuss potential
implications of these effects for the host galaxies: (i) radial mixing of
bulge stars with the rest of the host; (ii) contribution of quasar outflows to
galactic fountains as sources of high-velocity clouds; (iii) wholesale
ejection of hyper velocity stars out of their hosts, giving rise to type II
supernavae on galactic outskirts, and contributing to reionisation and metal
enrichment of the Universe; (iv) bulge erosion and even complete destruction
in extreme cases resulting in overweight or bulgeless SMBHs.
\end{abstract}


\section{Introduction}

Super-massive black holes (SMBH) accreting gas at rates comparable to the
Eddington accretion rate are believed to power quasars. They are expected to
launch powerful winds \citep{Shakura73,King03,Proga03c}. Such winds are
consistent with the fast $v_{\rm out} \sim 0.1 c$ winds detected via
absorption \citep[e.g.,][]{PoundsEtal03b,PoundsEtal03a} and also recently in
emission \citep{PoundsVaughan11a} in AGN X-ray spectra. These nuclear (most
likely $R \simlt 1$~pc)  winds must be wide-angle to explain their
detection frequency \citep{TombesiEtal10,Tombesi2010ApJ}, and energetically
should be capable of driving outflows clearing significant fractions of
{\em all} gas of the parent galaxy out \citep{SilkRees98,King05,ZK12a}. Recent
observations detect  such kpc-scale neutral and ionized outflows with
outflow velocities of $\sim 1000$ km~s$^{-1}$ and mass outflow rates of
hundreds to thousands of $\msun$~yr$^{-1}$
\citep[e.g.,][]{FergulioEtal10,SturmEtal11b,RP11a}, which are best interpreted
as the mass-loaded nuclear outflows. These outflows have wide-ranging
implications for galaxy formation and evolution \citep{DiMatteo05}.

Recently, \citet[hereafter NZ12]{NZ12} have shown that quasar outflows affect
the ambient gas in the bulge of the host galaxy in two different ways
depending on whether the ambient shocked gas is able to cool rapidly or
not. In the gas-poor epoch, e.g., the present day bulge of the Milky Way, the
shocked gas cools slowly, so the shock is approximately adiabatic. The layer
of the shocked gas is then geometrically thick and the density in it is $\sim
4$ times higher than the pre-shock density, as in the strong-shock limit. In
contrast, in the gas-rich epoch, when the pre-shock gas density in the bulge
is higher, the shocked gas cools more rapidly than the shock can propagate
through the bulge. The shocked gas is then swept up into a geometrically thin
shell, and the density in the shell is roughly $v_{\rm e}^2/c_{\rm s}^2$ times
the pre-shock density, where $v_{\rm e} \sim 1000$ km~s$^{-1}$ and $c_{\rm
  s}\simlt 10$ km~s$^{-1}$ are the shock front velocity and the sound speed in
the shell, respectively. If the pre-shock gas density is a $\simgt 0.1$
fraction of the total density responsible for the potential, the swept up
shell density is orders of magnitude higher than the tidal
density. Unsurprisingly then, NZ12 found that the shell can fragment into
star-forming clumps.

Such positive AGN feedback has been explored before with regard to jet
interaction with clumpy interstellar medium
\citep{Gaibler2012MNRAS,Silk2012A&A}. These authors found that AGN jets can
also enhance star formation in their host galaxies. However, the details of
that process are markedly different from what consider here and so we do not
attempt to draw direct comparison with these previous studies.

NZ12 concentrated on conditions when the shell stalls and falls back on to the
SMBH, which physically corresponds to black hole masses below the $M_\sigma$
mass \citep{King03,King05}
\begin{equation}\label{eq:msigma}
M_\sigma \simeq 3.67 \times 10^8 f \sigma_{\rm 200}^4 \; \msun,
\end{equation}
where $\sigma_{200}$ is the velocity dispersion in the host galaxy bulge, in
units of $200$ km/s, and $f$ is the ratio of the gas density in the bulge to
the total density in the bulge, scaled to the cosmological ratio of
$0.16$. Here we probe the evolution of the outflow for $\mbh \simgt
M_\sigma$. We find that the instability leading to the shell fragmentation is
still effective when the shell is being driven outward at velocities as large
as $\simgt 1000$ km s$^{-1}$, leading to formation of self-gravitating dense
gas clumps, and eventually newly born stars, moving radially outward with
these high velocities. This result suggests that quasar outflows could not
only be expelling gas from the host galaxies but also making some stars by
strong compression of that material; since the material is streaming away at
high velocity the new stars are born with that velocity as well.

Below we present numerical simulations that exemplify these effects and also
consider their theoretical implications.

\section{Numerical Methods}\label{sec:methods}

The simulations presented here to demonstrate our main point are set up in the
manner identical to those presented in NZ12, so we only briefly overview the
setup. For simplicity, the host galaxy is modelled by a fixed singular
isothermal sphere potential softened in the central $10$ pc where the SMBH is
located. The ambient gas is initialised as a sphere with inner radius $R_{\rm
  in} = 200$~pc and outer radius $R_{\rm out} = 10 R_{\rm in}$, at rest, and
with the density profile given by $\rho_{\rm g}(R) = f_{\rm g} \rho_0(R) =
f_{\rm g} \sigma^2 /2\pi G R^2$; $\sigma = 141$~km~s$^{-1}$ is the velocity
dispersion of the potential and $f_{\rm g} = 0.16 f$ is the ratio of the gas
density to the underlying density of stars and dark matter; we choose $f=1$
for the simulations presented in this paper. We use $5 \times 10^5$ SPH
particles in each simulation, giving a mass resolution $M_{\rm res} \simeq 40
m_{\rm SPH} = 2 \times 10^5 \; \msun$, much better than in a typical
galaxy-wide AGN feedback simulations \citep[e.g. $M_{\rm res} \simeq 6 \times
  10^6 \; \msun$ in][and much larger than this in cosmological
  simulations]{Springeletal05}. Initial gas temperature is set such that the
sound speed is equal to $\sigma$. At time $t=0$, the quasar outflow is turned
on.  The critical $M_\sigma$ mass above which even a purely momentum-driven
SMBH outflow should expel the gas for this potential is $M_\sigma \approx
1.5\times 10^8 f \msun$ \citep{NayakshinPower10}. Quasar feedback is
implemented via ``virtual particles'' emitted isotropically by the SMBH
\citep{NayakshinEtal09a,NayakshinPower10,ZN12a}. The quasar momentum outflow
rate is fixed at $\ledd/c$ and the energy input into the gas is $= 0.05
\ledd$; we use a smooth transition between the momentum- and energy-driven
flow regimes, with the cooling radius \citep{ZK12b} $R_{\rm c} = 500$~pc. The
SMBH mass is fixed for the duration of a simulation; this is a reasonable
assumption, given that the timescales we are interested in are shorter than
the Salpeter time ($\approx 40$ Myrs). Physically this situation could be
realised by a small scale dense gas disc feeding the SMBH at close to its
Eddington rate \citep[small scale gas discs are much more resilient to quasar
  feedback than extended spherical distributions; see][]{NPK12a}.

Since we are modelling the inner parts of a single galaxy, we can afford a
higher numerical resolution, as already noted above. Therefore, we treat star
formation in a way similar\footnote{Although due to a much poorer mass
  resolution here we do not attempt to model detail of sub-Solar protostar
  evolution, such as potential mergers of "First Cores" \citep{Larson69} or
  gas accretion onto the sink particles. Such microphysics of star formation
  is beyond the numerical realm of this paper.} to that introduced in
\cite{NayakshinEtal07}. Our approach is different from the industry practice
\citep[e.g.,][]{Springeletal05,BoothSchaye09,DuboisEtAl12} in modelling star
formation in cosmological simulation, where (i) a threshold density for star
formation is introduced (ranging from $10^4$ to $1$ Hydrogen atom per cm$^3$,
typically); and (ii) once gas density exceeds the threshold stars are
introduced {\em stochastically} following the 3D form of the Schmidt-Kennicutt
law \citep{Kennicutt98}. This method for treating star formation is reasonable
on galaxy-wide scales, provided that the threshold density exceeds the mean
gas density by a large factor. However, the method will fail in the inner
parts of bulges where the tidal density, $\rho_{\rm t} = \sigma^2/(2\pi G R^2)
\simeq 10^{-22} R_{\rm kpc}^{-2}$~g~cm$^{-3} \simeq 100 R_{\rm
  kpc}^{-2}$~cm$^{-3}$ is a strong function of radius $R$ and may be higher
than the threshold density. We also note that in the standard prescription,
once the {\it mean} gas density exceeds the density threshold then stars are
introduced everywhere in this prescription, even in gas regions that are not
gravitationally self-bound, e.g., in the hot interclump medium.

Our method avoids these potential pitfalls. Firstly, we require that the gas
is self-bound before star formation is initiated. This is done by requiring
that the Jeans mass, $M_{\rm J} \simeq 2.6 \times 10^6 T_4^{3/2} n_4^{-1/2} \;
\msun$, exceeds the minimum resolvable mass of the simulation, i.e., that of
40 SPH particles, $m_{\rm res} = 2 \times 10^5 \; \msun$; here $T_4 =
T/(10^4$~K$)$ and $n_4 = n/(10^4$~cm$^{-3})$. This criterion stops spurious
star formation in dense but hot gas (if such is formed in a strong shock, for
example). Secondly, we require that the gas density should exceed the tidal
density by a large factor, $A=200$. The latter is a free parameter, but we
tested that the results change insignificantly as long as $A$ is large, by
running a set of identical tests in which $A$ was varied from $2$ to
200. Physically, this second condition prevents star formation in regions that
appear to be self-gravitating but are not dense enough to withstand galactic
or bulge tides.

We are therefore confident that our methods correctly capture the onset of
star formation, and do this better than the industry standard. However, due to
our focus on the effects of AGN feedback, we neglect star formation feedback
within the gas clumps. The result of this approximation is that once star
formation occurs within the clump it is likely to consume all the available
gas, resulting in up to $100\%$ efficiency of star formation. This is most
likely incorrect. Star formation feedback could disrupt the star forming
clumps or heat them up, limiting star formation efficiency (although we note
that the star forming phase of our simulations is quite short, from a few Myr
to $\sim 10$ Myr, so only the most massive stars would have been able to
produce supernovae quickly enough). Concluding this discussion, our methods
correctly predict when star formation occurs, but we probably strongly
over-estimate the amount of stars formed. We shall keep this conclusion in
mind and come back to it in section 3.

Here we present three simulations, M1, M2 and M5, where the number indicates
the mass of the SMBH in units of $10^8 \; \msun$. Therefore in the first
simulation, we would expect the shell to stall if it was driven predominantly
by the momentum, rather than energy, input from the quasar. In the other two
simulations, the shell should escape to infinity. We present the results
below, drawing particular attention to the dynamics of the self-gravitating
clumps formed in the three simulations.

\section{Results}\label{sec:results}

\begin{figure}
\centerline{\psfig{file=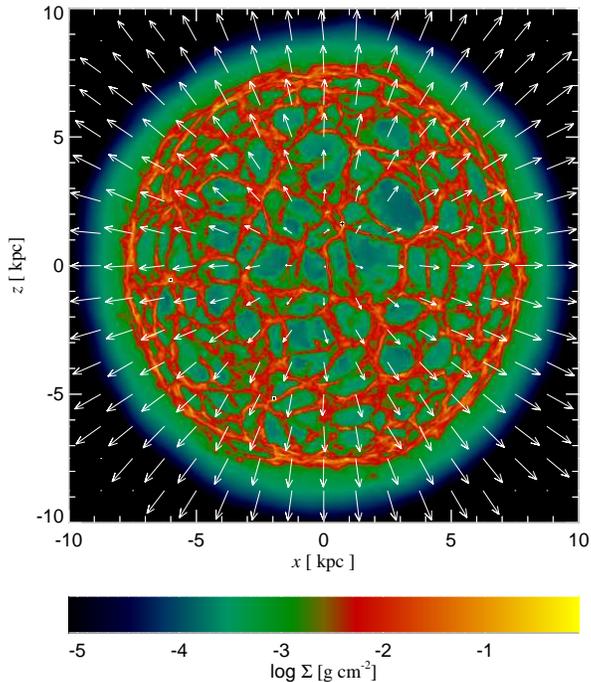,width=0.49\textwidth,angle=0}}
\caption{The gas column density and velocity map for simulation M5,
  11.8 Myr after the beginning of the simulation. Colour represents projected
  gas column density (scale at the bottom) and arrows indicate gas
  velocity. The arrows are logarithmically scaled to the longest one,
  so they only give qualitative information about gas motions.}
\label{fig:density}
\end{figure}

\begin{figure*}
\centerline{\psfig{file=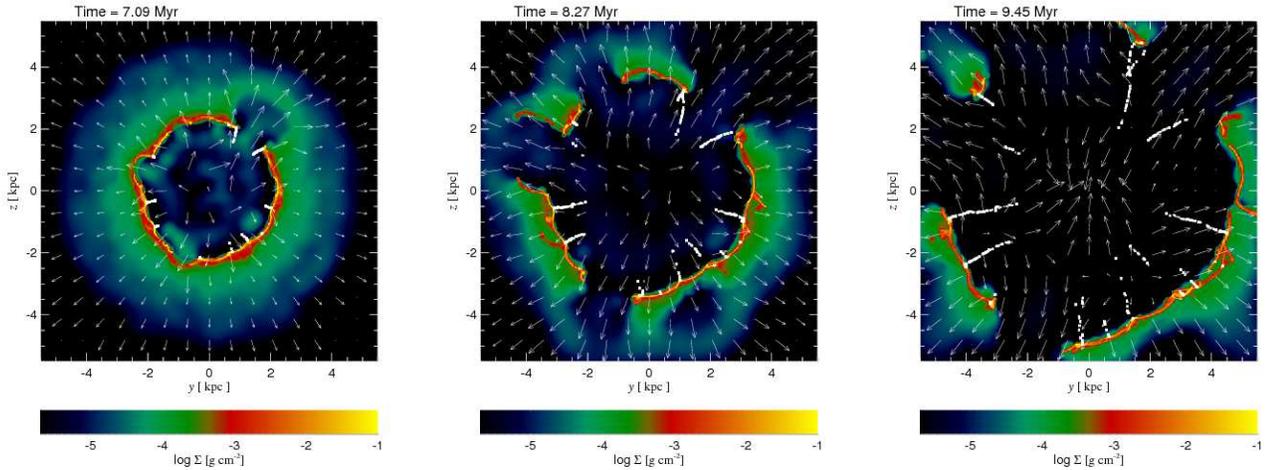,width=0.99\textwidth,angle=0}}
\caption{A wedge-slice through simulation M2 at times $7.09$,
  $8.27$ and $9.45$~Myr (left to right panels, respectively). Colours
  show projected gas surface density, white squares represent star
  particles. Stars form from overdensities in the fragmenting shell,
  but their clusters are stretched out because the stars decelerate
  while the shell moves outwards with approximately constant
  velocity. Arrows are gas velocity vectors, giving qualitative
  information about the turbulence in gas motion induced by quasar
  feedback.}
\label{fig:starform}
\end{figure*}

Figure \ref{fig:density} shows the column density for the simulation M5 and
velocity vectors projected on to the plane of the figure for gas at $11.8$ Myr
after the quasar switches on. We see that while the bulk motion of the gas is
radially outwards, there is a lot of structure forming as the shell
fragments. This is essentially the same result as found by NZ12: the cooling
time in the shocked ambient medium is $t_{\rm cool} \simeq 4 \times 10^5 v_8
R_{\rm kpc}^2$~yr, where $v_8 = v_{\rm e} / 1000$~km/s; this is a few times
shorter than the flow timescale $R \ v_{\rm e} \sim 10^6$~yr, so the outer
shell cools rapidly and is strongly compressed.  In this simulation, the
outflow does not stall, but rather keeps expanding and accelerating up to a
radial velocity $v_{\rm r} \simeq 2000$~km/s. This is because the SMBH mass is
above the critical $M_\sigma$ limit (eq. \ref{eq:msigma}) and so even gas
close to the SMBH is driven outside the cooling radius, where it can be
accelerated further by the energy input. The network of filaments produces
density enhancments that become self-gravitating.

\begin{figure}
\centerline{\psfig{file=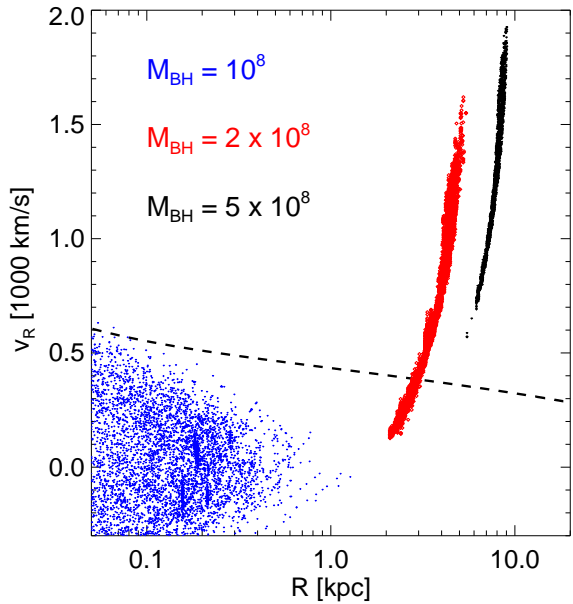,width=0.49\textwidth,angle=0}}
\caption{Radial velocity of stars formed in simulations M1, M2 and M5 versus
  their instanteneous radial coordinate, $R$. The dashed line shows the escape
  velocity as a function of $R$. The star forming shell created by the
  lightest black hole stalls at a radius of $R\sim 0.3$ kpc, so all of the
  stars formed in this simulation are bound to the host. }
\label{fig:fig2}
\end{figure}

In Figure \ref{fig:starform}, we show a narrow wedge projection (see NZ12 for
more detail) of the simulation M2 at $7.09$, $8.27$ and $9.45$~Myr. White
squares show star particles. Note that they form filaments pointing mainly in
the radial direction, similar to the ``fingers of creation'' found in
NZ12. Denser gas regions experience a lower outward push per unit mass, and
therefore lag slighly behind lower density gas. These dense filaments
eventually collapse into self-gravitating clumps, which are then observed to
lag behind the shell. Clumps born earlier on are born with lower radial
velocity since the gas shell is being accelerated outward due to its
decreasing mean projected column density. Therefore dense gas regions
collapsing over the course of the simulations result in ``streamers'' of star
particles clearly visible in Figure 2. For the simulation M2 in particular,
the first clumps move with mean radial velocities $v_* \sim 200$~km/s,
slightly lower than the bulge escape velocity $v_{\rm esc} \simeq 400$~km/s
(see also red points in Figure \ref{fig:fig2}).

Figure \ref{fig:fig2} shows the radial velocity versus radius for star
particles formed in all three simulations. The snapshots are taken at
$16.5$~Myr for M1 and at $9.45$~Myr for the other two simulations. The dashed
line in the simulation shows the escape velocity as a function of radius,
assuming the effective radius for the model galaxy of $20$ kpc. In simulation
M1, the shell fragments, the dense clumps stall and then fall back on to the
SMBH. For this reason, there are both positive and negative radial velocities
for stars in this simulation. Almost all of the stars formed in this
simulation (blue dots) are bound to the host galaxy.

For the more massive SMBHs, the shell is continuously driven out. As the shell
is of a finite mass, its mean projected column density decreases as it
expands, and so quasar outflow can accelerate it to velocities larger than
would be calculated analytically for a shell of infinite radius \citep[see,
  e.g.,][for derivation]{KZP11}. As already explained above, the collapsing
star-forming clumps do not experience the same outward acceleration from the
quasar outflow and decouple from the shell. Star particles born earlier are
born with a lower outward velocity, therefore they lag behind those that are
born at later time. As a result the dispersion in clump velocities increases
with time, which is already evident from Figure \ref{fig:starform}. Figure
\ref{fig:fig2} shows that some of the stars formed in simulation M2 are bound
to the galaxy (those below the dashed curve); they move on very elongated,
nearly radial orbits. Stars formed later have large enough radial velocities
to escape the potential completely. In simulation M5 the whole of the shell is
accelerated to velocities above the escape velocity before the stars are born;
all of the newly made stars are on escaping trajectories.

A large fraction, $30-70\%$, of the initial mass of the shell is converted
into star particles in each of the three simulations. As discussed in section
\ref{sec:methods}, such a large star formation efficiency is clearly an
overestimate. Star formation feedback, neglected here, is likely to reduce the
total mass of the newly made stars. Nevertheless, even if star formation
efficiency drops by an order of magnitude, e.g., to $5\%$, the total mass of
stars formed on the radial outward trajectories would be between a few times
$10^7 \; \msun$ to $10^8\msun$. We also note that star formation efficiency is
expected to be higher in compressed systems \citep{Keto2005ApJ}. Thus,
provisionally, we may expect at least $\sim 10^8 \; \msun$ of stars to be
formed.

\section{Discussion}

The simulations presented here show that, in the setup explored, quasar
feedback can expel gas from the bulge of its host galaxy with high velocity
and simultaneously compress it to high densities. These two processes result
in formation of hyper-velocity stars (HVS). The total mass of these stars is
unclear at the moment, since our simulations neglect star formation feedback
(see \S \ref{sec:methods} and \ref{sec:results}). In addition, here we
explored the simplest possible initial conditions -- a spherically symmetric
gas distribution, with the density profile $\rho \propto 1/R^2$. Simulations
of AGN feedback with more realistic non-spherical initial conditions show that
the quasar outflow driven bubble may burst in directions of the lowest column
depth of the ambient gas, e.g., perpendicular to the planet of the host galaxy
\citep{NZ12}. Feedback in such a geometry may be \citep{ZN12a} the reason
behind formation of the "Fermi Bubbles" in the Milky Way \citep{SuEtal10}. In
such geometries the fraction of ambient gas put on hyper-velocity orbits is
smaller, and therefore we expect that the total mass of the HVS would be also
smaller.

Therefore, while we are confident that AGN feedback can launch HVS, we are
currently unable to quantify how often this occurs in more realistic
situations, and how large the fraction of gas turned into stars is. With this
caviat in mind, we now discuss possible observational implications of HVS made
by AGN feedback.

\subsection{Radial mixing of stars and clouds}

Stars formed in the shock driven outward by the quasar with velocities below
the escape velocity, such as those shown with the red symbols that are below
the dashed curve in Figure \ref{fig:fig2}, are destined to fall back to the
bulge. These stars are on very elongated nearly radial trajectories, and spend
most of the time at large radii, i.e., $R\simgt$ a few to a few tens of
kpc. These stars are thus unusual in that they were born from the gas
initially filling the inner region of the galaxy, the bulge, but due to their
large radial velocity kicks now belong to the galactic halo. Such halo stars
could thus stand out due to their highly radial orbits and (likely) higher
metallicity compared to a typical low metallicity halo star. The most extreme
of these stars, in the outskirts of the dark matter halos of their host
galaxies, may provide a significant contribution to the intergalactic infrared
background. It was recently shown that a $0.1 \%$ fraction of total galaxy
light deposited into the outskirts of halos could explain most of the
background radiation \citep{Cooray2012Natur}; while we cannot quantify the
total luminosity of stars expelled to such distances in our model, our results
are consistent with this requirement.

The gas clumps expelled from the bulge do not necessarily all form stars. Some
of these clumps may remain marginally bound, and collapse on a timescale
longer than the cloud ejection timescale. In that case, the gas clumps can mix
with the halo gas, joining the high velocity cloud (HVC) population. These
clouds are observed all around the Milky Way, have high peculiar velocities
and metallicities similar to or slightly lower than Solar
\citep{Wakker1997ARA&A,Wakker2001ApJS}. Their evolutionary timescales are
estimated at a few tens to several hundred Myr, suggesting continuous
replenishment, such as condensation from tidal streams \citep{Putman2003ApJ}
or a galactic fountain \citep{Bregman1980ApJ,Kwak2009ApJ}. Our simulations
suggest that quasar outflow, long extinguished by now,  could also
contribute to production of HVCs.

\subsection{Bulge disruption}

It is usually assumed that gas is a small fraction of the total mass of the
host, with the rest consisting of dark matter and stars. This is clearly
correct for present day galaxies, and must be globally true for galaxies at
birth since the universal baryon fraction is only $f_{\rm g} \sim
0.16$. However, gas can collapse to the centre of the potential if its
  angular momentum is sufficiently low and so dominate the potential in the
central region of the host. In this case, if most of the gas is
driven rapidly out of the host, the remaining mass may be insuffient to bind
the bulge gravitationally. It then has to expand and perhaps blend with the
rest of the galaxy, adding stars to the halo, with faster stars leaving the
host galaxy altogether.

In the extreme cases this should result in an SMBH without a clear parent
bulge or an SMBH that appears to be far more massive than it should be based
on the $\mbh-M_{\rm bulge}$ relation. This cannot be a common outcome: based
on Figure 3, we see that the SMBH mass needs to be somewhat above the
$M_\sigma$ value to drive the cold shell out before the latter collapses into
stars. This happens even in a spherically-symmetric geometry, which is more
conducive to large outflows; NZ12 found that for more realistic non-spherical
geometries, AGN feedback in the gas rich epoch tends to compress and trigger
star bursts in cold gas without driving the shell away.

Nevertheless, bulge destruction or erosion (partial destruction) by the SMBH
feedback expelling a large fraction of the bulge mass may occasionally happen,
especially for active galaxies in the early Universe and/or violently merging
systems. This may be relevant for the recently observed population of SMBHs in
bulgeless galaxies \citep{Simmons2012arXiv,McAlpine2011ApJ,ArayaSalvo2012} and
well as several SMBHs with masses significantly above those predicted by the
black hole - bulge mass correlation \citep{Bogdan2012ApJ}.

\subsection{Type II supernovae on galactic outskirts}

The distance travelled by a star moving with radial velocity $v = 1000 v_8$ km
s$^{-1}$ for time $t = 10^7 t_7$ yrs is
\begin{equation}
\Delta R = v t \approx 10 \hbox{ kpc } v_8 t_7\;.
\label{deltar}
\end{equation}
Stars with Solar or super-Solar metallicity that are more massive than $M\sim
8\msun$ end their main sequence life in a Type II supernova explosion
\citep{Heger2003ApJ}. The main sequence lifetime of stars of mass $M=8\msun$
is approximately 100 Myr, while the most massive stars, $M\simgt 50 \msun$
live only $\sim 4$ Myr. Therefore we can expect that massive stars launched by
the quasar outflow produce Type II supernovae at a distance from a few to few
tens of kpc away from the galaxy's center. Additionally, long duration GRBs
are expected from this flung-out population of stars \citep{Woosley2006ARA&A},
potentially explaining the GRBs that are known to be offset, some by as much
as $> 10$~kpc, from the centres of their host galaxies \citep{Bloom2002AJ}.

\subsection{Reionisation and metal enrichment of the Universe}

\cite{BouwensEtal11a} show that stars formed in galaxies at around the
reionization epoch can provide sufficient amount of ionizing photons provided
that they escape into intergalactic space efficiently enough. In particular,
these authors estimate that an escape fraction of 20\% is required for
consistency with the WMAP results at the $2\sigma$ level. This is much higher
than the escape fraction of only $\sim 1 - 2$\% as found for the Milky Way
\citep{PutmanEtal03a}. The disagreement is all the more striking because the
present day Milky Way contains very little gas per unit total mass of the
Galaxy compared with the early star forming galaxies, which were also probably
much more compact given that observations show that galaxies grew inside out
at least from $z\sim 2 $ to the present day \citep{vanDokkumEtal10}.

It is therefore interesting to note that the massive stars launched at high
velocities, $v \simgt v_{\rm esc}$, find themselves outside the cold high
density gas environment of the inner galaxy. Ionizing photons emitted by these
stars have orders of magnitude smaller average gas column depth to navigate
through to escape from the galaxy's halo. Note that while massive stars are
trailing the cold shell that created them earlier on, the shell is a web of
thin dense filaments with most of the sky covered by a much lower density
material (see Fig. \ref{fig:density} and also Fig. 2 in NZ12). Therefore, the
escape fraction of ionising radiation for these stars may be much closer to
unity than for the stars residing in the inner galaxy. The Hyper Velocity
Stars (HVS) formed during quasar outflow ejection may thus be far more
efficient in terms of reionization than the ``normal'' galactic stars,
although it remains to be seen if a sufficient mass of HVS can be created to
contribute significantly to the reionization of the Universe.

Lower mass stars ejected of their host galaxies at high velocities may be
contributing to metal enrichment of the gas outside of galaxies. Their longer
main sequence lifetime, $t\sim 10^9$ yrs, implies that they may travel as far
as $\sim 1$ Mpc (see eq. \ref{deltar}) from their hosts before ejecting their
envelopes. Recent observations have discovered low-mass stars at distances of
up to $2$~Mpc from the centre of the Milky Way \citep{Palladino2012AJ}. These
stars might have been ejected by three-body interactions between a binary star
and the central SMBH of the Milky Way \citep[e.g.,][]{Yu2003ApJ}, but without
knowledge of their orbits, it is impossible to confirm or disprove this
scenario. The existence of these stars is also consistent with our model; they
might have been ejected by a quasar outflow from the Milky Way several Gyr
ago. Importantly, this ``stellar outflow'' may actually reach further into the
intergalactic space than quasar gas outflows. Here we found that stars are
lagging behind gas outflows while quasar outflow provides an outward
acceleration for the gas shell; but eventually the quasar must turn off, and
the gas outflow must decelerate faster than a HVS because the outflow
interacts with intergalactic medium by a shock whereas the stars continue to
travel ballistically.

\section{Summary and conclusion}

Using numerical simulations of SMBH feedback upon gas in its host galaxy, we
have shown that quasars may not only trigger star formation in their host
bulges \citep[as shown by][]{NZ12}, but also expel a fraction of these
newly-formed stars from the host galaxy. Expulsion requires a large SMBH mass
($M > M_\sigma$), so may not be a common phenomenon. Further modelling is
needed to determine the mass fraction produced in this HVS population, since
its potential implications are wide ranging.

Stars originating in the bulge launched at high velocities on nearly radial
trajectories may completely escape the galaxy; slower stars add a metal-rich
subpopulation in the region and may contribute to the intergalactic
light. High metallicity HVS produce ionising radiation and so may contribute
to the reionisation of the Universe; they also explode as supernovae,
potentially contributing to Type II supernovae on galactic outskirts. Lower
mass stars travel further out, enriching the extragalactic medium with
metals. Finally, a significant fraction of the galaxy bulge mass may become
unbound due to such ``stellar outflows'', leading to SMBHs in bulgeless
galaxies or SMBHs with masses much higher than expected from the black hole -
bulge mass relation. We conclude that quasar outflows shape the properties of
their galaxies and surrounding medium in many ways, not just by removing gas
and quenching star formation, as usually acknowledged.

\section*{Aknowledgments}

This research used the ALICE High Performance Computing Facility at the
University of Leicester.  Some resources on ALICE form part of the DiRAC
Facility jointly funded by STFC and the Large Facilities Capital Fund of BIS.
Theoretical astrophysics research in Leicester is supported by an STFC Rolling
Grant.

\label{lastpage}

\end{document}